\documentclass[twocolumn,aps]{revtex4-1}

\usepackage{graphicx}
\usepackage{amssymb}
\usepackage{amsmath}
\usepackage{epsfig}
\usepackage{setspace}
\usepackage[caption=false]{subfig}
\usepackage{ulem}
\usepackage{bm}

\begin{document}

\title{Diffusive dynamics of critical fluctuations near the QCD critical point}

\author{Marlene Nahrgang$^{1}$, Marcus Bluhm$^2$, Thomas Sch\"afer$^3$, Steffen A. Bass$^4$}

\affiliation{$^1$SUBATECH UMR 6457 (IMT Atlantique, Universit\'e de Nantes, \\IN2P3/CNRS), 4 rue Alfred Kastler, 
44307 Nantes, France}
\email{marlene.nahrgang@subatech.in2p3.fr}

\affiliation{$^2$Institute of Theoretical Physics, University of Wroclaw, PL-50-204 Wroclaw, Poland}

\affiliation{$^3$Physics Department, North Carolina State University, Raleigh, NC 27695, USA}

\affiliation{$^4$Department of Physics, Duke University, Durham, NC 27708-0305, USA}

\begin{abstract}
A quantitatively reliable theoretical description of the dynamics of fluctuations in non-equilibrium is indispensable 
in the experimental search for the QCD critical point by means of ultra-relativistic heavy-ion collisions. In this 
work we consider the fluctuations of the net-baryon density which becomes the slow, critical mode near the critical 
point. Due to net-baryon number conservation the dynamics is described by the fluid dynamical diffusion equation, 
which we extend to contain a white noise stochastic current. Including nonlinear couplings from the 3d Ising model 
universality class in the free energy functional, we solve the fully interacting theory in a finite size system. 
We observe that purely Gaussian white noise generates non-Gaussian fluctuations, but finite size effects and 
exact net-baryon number conservation lead to significant deviations from the expected behavior in equilibrated 
systems. In particular the skewness shows a qualitative deviation from infinite volume expectations. With 
this benchmark established we study the real-time dynamics of the fluctuations. We recover the expected dynamical 
scaling behavior and observe retardation effects and the impact of critical slowing down near the pseudo-critical 
temperature. 
\end{abstract}

\maketitle

\section{Introduction}
\label{sec:Intro}

In local thermal equilibrium the physics of long-wavelength phenomena can be described effectively by fluid dynamics. 
Conventional fluid dynamics specifies how energy and momentum as well as charges such as the net-baryon 
number are conserved on average. However, in reality fluids experience noise and dissipation~\cite{Kovtun:2012rj}. 
Intrinsic thermal fluctuations in the fluid dynamical quantities drive the system constantly out of equilibrium. 
Dissipative processes balance the impact of the fluctuations such that local equilibrium emerges as a consequence of 
this balance. Near phase transitions, in particular near a critical point, the influence of fluctuations is drastically 
enhanced and cannot be considered as a small perturbation. Thus, a direct and full propagation of fluctuations in the 
fluid dynamical evolution is essential to understand fluctuation-driven dynamical effects in systems at a critical 
point. 

The search for the critical point in the QCD phase diagram has attracted special attention in recent years. It is at 
the heart of many heavy-ion collision experiments performed at GSI/SIS18, the CERN-SPS and in the beam energy scan at 
RHIC. As a signature of its existence a non-monotonic behavior in observables related to event-by-event fluctuations 
of conserved charges is expected with varying beam energy~\cite{Stephanov:1998dy}. Indeed, recent data of net-proton 
number fluctuations reveal signs of non-monoticity~\cite{Adamczyk:2013dal,Thader:2016gpa}. The global picture, however, 
remains inconclusive, mostly due to the lack of quantitative theoretical predictions.

Near a critical point, fluctuations of the critical mode grow as they scale with the correlation length $\xi$. 
Importantly, higher-order cumulants are sensitive to higher powers of $\xi$~\cite{Stephanov:2008qz,Asakawa:2009aj}.
The slow, critical mode associated with the divergence of the correlation length at the QCD critical point is the 
net-baryon density $n_B$~\cite{Son:2004iv}. Enhanced fluctuations in the conserved net-baryon number are therefore 
expected if the matter passes through the critical region. 

These expectations are based on the assumption that the critical mode is in equilibrium with the rest of the matter. 
However, the system created in a heavy-ion collision is small and short-lived, and due to its violent expansion dynamics 
spends only a short time in the critical region. Thus, non-equilibrium effects are 
essential~\cite{Berdnikov:1999ph,Nahrgang:2011mg,Mukherjee:2015swa,Herold:2016uvv,Sakaida:2017rtj,Stephanov:2017ghc}. 
The time for $\xi$ to reach the equilibrium value scales as $\xi^z$. For a purely diffusive dynamics near the critical 
point, $z\simeq 4$ according to the dynamical universality class of model B~\cite{Hohenberg:1977ym}. Thus the actual 
growth of $\xi$ is already dynamically limited by the lifetime of the system~\cite{Berdnikov:1999ph}. 
In the search for the QCD critical point, the development of dynamical models which capture the non-equilibrium evolution 
is therefore crucial. This work reports first, important results from such a dynamical model.

\section{Diffusive dynamics of critical fluctuations}
\label{sec:Theory}
The evolution of the regular bulk matter created in a heavy-ion collision can successfully be described by conventional, 
relativistic fluid dynamics with a surprisingly small shear viscosity over entropy density ratio, 
see~\cite{Romatschke:2017ejr} for a recent review. Starting from the equations of relativistic fluid dynamics 
describing the conservation of charges, energy and momentum 
\begin{equation}
 \partial_\mu N^\mu = 0\, , \quad\quad
 \partial_\mu T^{\mu\nu} = 0\, ,
\end{equation}
we focus in this work on the conservative evolution of the net-baryon number $N_B$ decoupled from the fields of energy and 
momentum density. The diffusive dynamics occurs such as to minimize the free energy of the system. Since we are interested 
in the dynamics of intrinsic fluctuations near the QCD critical point we include a stochastic current and obtain the 
following stochastic partial differential equation (similar to~\cite{Hohenberg:1977ym,delaTorre:2014mys}) 
\begin{equation}
 \label{equ:diffeq}
 \partial_t n_B({\bf x},t) = \Gamma \nabla^2\left({\cal F}'[n_B]\right) + \nabla\cdot{\bf J}({\bf x},t) \,.
\end{equation}
Here, $\Gamma$ is the mobility coefficient, ${\cal F}[n_B]$ denotes the free energy functional, ${\cal F}'$ its variation 
and ${\bf J}({\bf x},t)$ is the stochastic current given by 
\begin{equation}
 \label{equ:stochcurrent}
 {\bf J}({\bf x},t) = \sqrt{2T\Gamma} {\bm{\zeta}}({\bf x},t) 
\end{equation}
for temperature $T$. The spatio-temporal white noise field ${\bm{\zeta}}({\bf x},t)$ is Gaussian with zero mean and 
covariance $\langle\zeta_i({\bf x},t)\zeta_j({\bf x'},t')\rangle=\delta({\bf x}-{\bf x'})\delta(t-t')\delta_{ij}$. 
Eq.~(\ref{equ:stochcurrent}) guarantees that the fluctuation-dissipation balance is fulfilled and the equilibrium 
distribution $P_{\rm eq}[n_B]$ is given by the known statistical expression. 

Near the critical point we approximate the free energy functional in Ginzburg-Landau form as 
\begin{multline}
 \label{equ:freeenergy}
 {\cal F}[n_B] = T\int{\rm d}^3 x \left(\frac{m^2}{2n_c^2}\left(\Delta n_B\right)^2 + \frac{K}{2n_c^2}
 (\nabla n_B)^2\right. \\
 \left. + \frac{\lambda_3}{3n_c^3}\left(\Delta n_B\right)^3 + \frac{\lambda_4}{4n_c^4}\left(\Delta n_B\right)^4 + 
 \frac{\lambda_6}{6n_c^6}\left(\Delta n_B\right)^6 \right) 
\end{multline}
with $\Delta n_B=n_B-n_c$ and $n_c$ is the critical net-baryon density. Apart from the typical Gaussian mass term which 
gives rise to the standard diffusion equation, Eq.~(\ref{equ:freeenergy}) contains a kinetic term modified by the surface 
tension $K$ as a measure for the range of the interaction as well as nonlinear interaction terms. 

To make contact with the static universality class of QCD we define $K=\tilde{K}/\xi_0$, $m^2=\tilde m^2/\xi_0^3$, 
$\lambda_3=n_c\tilde{\lambda}_3\tilde m^{3/2}$, $\lambda_4=n_c\tilde{\lambda}_4\tilde m$ and $\lambda_6=n_c\tilde{\lambda}_6$. 
Here, the dimensionless mass $\tilde{m}$ vanishes at the critical point and can be associated with the equilibrium correlation 
length in the noninteracting Gaussian limit. Finally, $\xi_0=0.48$~fm is a characteristic length scale of the system far away 
from criticality, e.g. the scale of noncritical correlations. These definitions ensure that the coupling coefficients exhibit 
the universal critical scaling with $\tilde m$~\cite{Tsypin:1994nh,Tsypin:1997zz} known from the 3-dimensional Ising model. In 
contrast to previous considerations we include a term proportional to $\left(\Delta n_B\right)^6$ in the free energy functional, 
Eq.~(\ref{equ:freeenergy}), because this turned out to be important for a quantitative description of lattice Monte Carlo 
results for the probability distribution of the average magnetization in the Ising model~\cite{Tsypin:1994nh,Tsypin:1997zz}. We 
thus obtain the nonlinear stochastic diffusion equation for a system with homogeneous temperature 
\begin{multline}
 \partial_t n_B({\bf x},t) = \frac{D}{n_c} \left(m^2 \nabla^2 n_B - K\nabla^4 n_B\right)\\ 
 + D\nabla^2\left(\frac{\lambda_3}{n_c^2}\, \left(\Delta n_B\right)^2 + \frac{\lambda_4}{n_c^3}\, \left(\Delta n_B\right)^3 + 
 \frac{\lambda_6}{n_c^5}\, \left(\Delta n_B\right)^5\right)\\ + \sqrt{2Dn_c}\nabla\cdot\bm{\zeta}({\bf x},t)
 \label{equ:stochdiff}
\end{multline}
with the diffusion coefficient $D=\Gamma T/n_c$. 

The exact value of the critical net-baryon density is sensitive to the location of the critical point in the QCD phase diagram 
and to the equation of state. We assume $n_c=1/3$~fm$^{-3}$, which we also choose as the average value of $n_B$ independent of 
$T$. On the crossover side of the critical point we consider the temperature dependence of $\xi_0/\tilde m$ as shown in 
Fig.~\ref{fig:integratedmoments} (black solid line in the lower inlay). 
This is determined by a matching to the susceptibility of the Ising model scaling equation of state~\cite{Guida:1996ep} near 
the critical point. The dimensionless mass $\tilde m$ is minimal near $T_c=0.15$~GeV and approaches $1$ for $T\to 0.5$~GeV. 
Values for the dimensionless universal couplings $\tilde\lambda_3$, $\tilde\lambda_4$ and $\tilde\lambda_6$ can similarly be 
deduced from the scaling equation of state, see~\cite{Bluhm:2016trm}. Here, we consider the constant values $\tilde\lambda_3=1$, 
$\tilde\lambda_4=10$ and $\tilde\lambda_6=3$ which are also comparable with the lattice Ising model studies 
in~\cite{Tsypin:1994nh,Tsypin:1997zz}. The values are such that near $T_c$ the nonlinear couplings $\lambda_4$ and $\lambda_6$ 
are large compared to $\tilde m^2$ and, thus, interesting nonlinear effects can be expected in this temperature region. 
In the following we use $\tilde K=1$, for which the simple relation $\xi=\xi_0/\tilde{m}$ is recovered in the 
Gaussian limit~\cite{Bluhm:2018plm}. 
For all our results in the remainder of this work we consider fully dynamical fluctuations in one spatial dimension. In the 
context of a heavy-ion collision this means that we treat fluctuations in the transverse direction as independent Gaussian 
variables, but treat fluctuations in the longitudinal direction using the full nonlinear theory. 

\section{Fluctuation observables in equilibrium}

\begin{figure}
 \centering
 \includegraphics[width=0.48\textwidth]{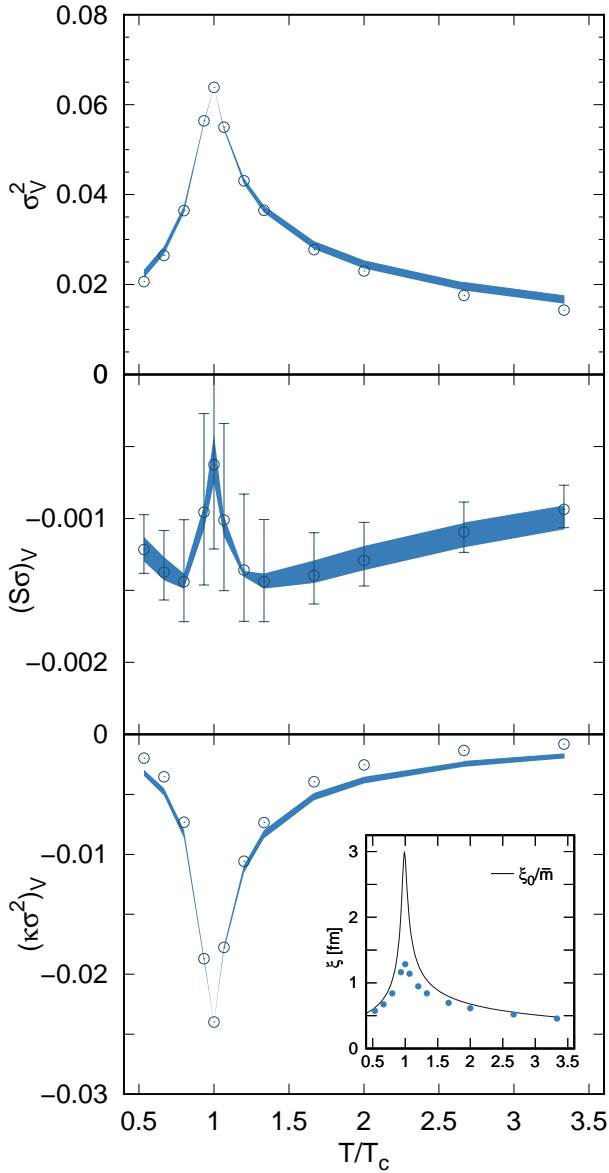}
 \caption{Variance $\sigma^2_V$, skewness $(S\sigma)_V$ and kurtosis $(\kappa\sigma^2)_V$ of the net-baryon number (open 
 circles) in the observation region $V\simeq1.95$~fm as a function of scaled temperature $T/T_c$. For variance and kurtosis 
 the statistical errors are smaller than the symbols. The colored bands demonstrate the scaling behavior with the equilibrium 
 correlation length $\xi$ as discussed in the text. The inlay in the lower panel shows $\xi$ for the fully interacting theory 
 (circles) and for the Gaussian limit (solid line).}
 \label{fig:integratedmoments}
\end{figure}
Let us first evaluate the fluctuation observables in an equilibrated system and thus establish a benchmark for the dynamical 
calculations. At each temperature we fix $D=1$~fm and take the results from the late-time limit. 
The equilibrium correlation length $\xi$ of the fully interacting theory can be read off from the exponential decay behavior 
of the equal-time spatial correlator $\langle \Delta n_B(r)\Delta n_B(0)\rangle$. The result is shown in the inlay of 
Fig.~\ref{fig:integratedmoments} (blue circles). 
The impact of the nonlinear interaction terms and finite size effects manifests itself as a significant reduction of the 
correlation length near $T_c$ compared to the Gaussian limit. 

For various temperatures we calculate the equilibrium values of the fluctuation observables of the net-baryon number in a 
given volume, i.e.~the variance $\sigma^2_V$, the skewness $(S\sigma)_V$ and the kurtosis $(\kappa\sigma^2)_V$, in the fully 
interacting theory. Here, $\langle N_B\rangle_V =\int_{V} n_B(x) {\rm d} x$ can take continuous values in each event, and 
$(S\sigma^3)_V$ and $(\kappa\sigma^4)_V$ denote the third- and fourth-order cumulants of the event-by-event distribution of 
$\langle N_B\rangle_V$. The volume $V$ corresponds to a  subregion of observation smaller than the system size, $V<L=20$~fm. 
This is analogous to the experimental situation in which $N_B$ is measured in a given kinematic region, e.~g.~in a given 
rapidity window for heavy-ion collision experiments. Our lattice spacing is $\Delta x=L/256$. 

In Fig.~\ref{fig:integratedmoments} we show our results for an observation volume $V\simeq 1.95$~fm (open symbols). In all 
three quantities, we clearly observe a prominent signal around $T_c$, where the equilibrium correlation length is the largest. 
Due to the nonlinear interactions the non-Gaussian fluctuations $(S\sigma)_V$ and $(\kappa\sigma^2)_V$ are created from purely 
Gaussian white noise in Eq. (\ref{equ:stochdiff}). The colored bands show fits for the temperature range near $T_c$ measuring 
the scaling behavior of the fluctuation observables with $\xi$. We find an optimal description for 
$\sigma_V^2\propto\xi^{1.30\pm 0.05}$ and $(\kappa\sigma^2)_V\propto\xi^{2.5\pm 0.1}$. These results demonstrate a clear 
reduction of the scaling behavior compared to the leading-order (assuming $\xi/V\ll 1$) expectations $\sigma_V^2\sim\xi^2$ and 
$(\kappa\sigma^2)_V\sim\xi^5$ in~\cite{Stephanov:2008qz}. 
Solving the fully interacting theory for a finite size system leads to a complicated interplay between nonlinear mode 
couplings, finite $\xi/V$ effects and exact net-baryon number conservation. We expect, for example, that the nonlinear mode 
couplings $\lambda_3\lambda_4$ and $\lambda_3\lambda_6$ can affect the skewness near $T_c$. For $\xi/V<1$ this can be studied 
systematically~\cite{WorkInProgress}. In Fig.~\ref{fig:integratedmoments} we observe indeed a competition between different scaling 
behaviors in the skewness whose infinite volume expectation~\cite{Stephanov:2008qz} is $(S\sigma)_V\sim\xi^{2.5}$. 
We find instead an optimal description with a term $\propto\xi^{1.47\pm 0.05}$ and a competing contribution of opposite 
sign $\propto\xi^{2.40\pm 0.05}$. 
We conclude that already in equilibrium for the fully interacting theory and including net-baryon number conservation in a 
finite size system the scaling behavior is drastically different from what has been expected so far. 

\section{Dynamics of fluctuation observables}

\begin{figure}[tb]
 \centering
 \includegraphics[width=0.48\textwidth]{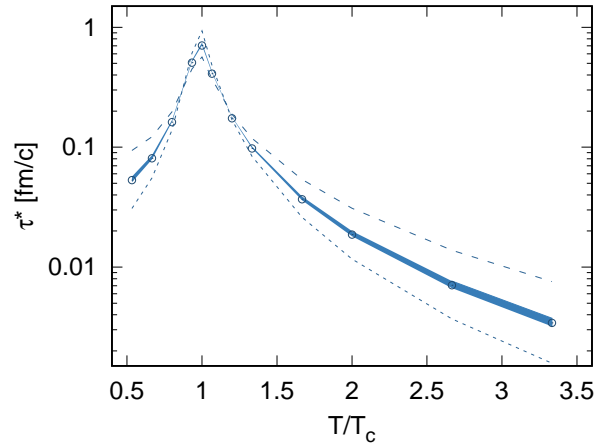}
 \caption{Scaling behavior of the relaxation time $\tau^*$ (circles) with $\xi$ for modes $k^*=1/\xi$ as a function of 
 $T/T_c$. The colored band shows the scaling $\propto\xi^z$ with $z=4\pm0.1$.} 
 \label{fig:RelaxationTimeScaling}
\end{figure}
Having settled the equilibrium benchmark for net-baryon fluctuations in a finite system, we can now turn to the dynamics of 
the critical fluctuations. In the following we consider a diffusion coefficient $D$ that depends on temperature as $D=D_0T/T_0$ 
with $D_0=1$~fm at $T_0=0.5$~GeV. As a first quantity we study the dynamical structure factor and the dynamical scaling behavior 
near $T_c$. For this purpose, we analyze the correlator $\langle\Delta n_B(k,t_0+t)\Delta n_B(-k,t_0)\rangle$ for different 
wavevectors $k$. It is found to decay over time $\propto e^{-t/\tau_k}$. As one would expect, the relaxation time $\tau_k$ 
decreases with increasing $k$ and becomes larger for fixed mode $k$ as $T\to T_c$. 

Considering the modes with $k^*=1/\xi$ for the correlation length realized in the fully interacting theory at a given $T$, we 
find that the corresponding relaxation time $\tau^*$ scales as $\tau^*\propto \xi^z$ with $z\simeq 4$. This is exhibited in 
Fig.~\ref{fig:RelaxationTimeScaling} where we also contrast the scaling behavior with the scaling exponents $z=3$ (dashed line) 
and $z=5$ (dotted line), which give a poor description and are clearly excluded. The excellent agreement with $z\simeq 4$ demonstrates 
that the expected dynamical critical scaling of model $B$ is reproduced. 
\begin{figure}[t]
 \centering
 \includegraphics[width=0.48\textwidth]{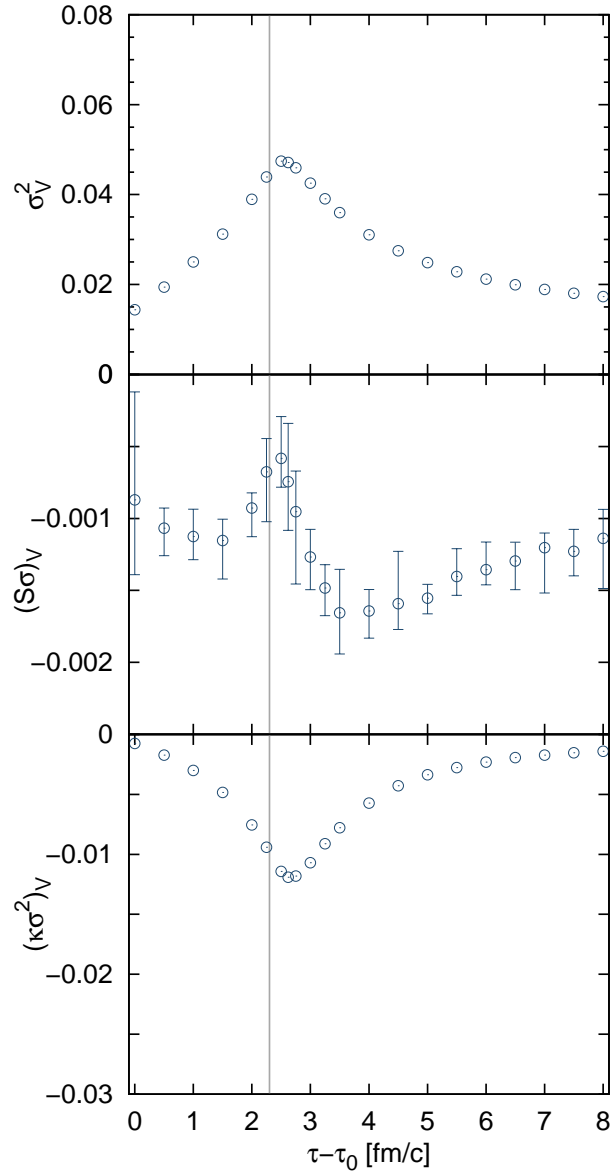}
 \caption{Dynamical evolution of the variance $\sigma^2_V$, the skewness $(S\sigma)_V$ and the kurtosis $(\kappa\sigma^2)_V$ as 
 a function of time. The pseudo-critical temperature $T_c$ is reached at $\tau_c=2.3$~fm/c.}
 \label{fig:integratedmomentsDyn}
\end{figure}

Next, we consider a dynamical evolution of the background temperature of the system according to
\begin{equation}
 T(\tau)=T_0\left(\frac{\tau_0}{\tau}\right)\, .
\end{equation}
Now, the coupling constants and the diffusion coefficient are time dependent. At $\tau_0=1$~fm/c we start with an equilibrated 
system at $T_0=0.5$~GeV. For the given temperature evolution $T_c$ is reached at around $\tau_c=\tau-\tau_0=2.3$~fm/c. 

In Fig.~\ref{fig:integratedmomentsDyn}, we present the results of the real-time dynamics for the volume-integrated variance, 
skewness and kurtosis. The time at which $T_c$ is reached is indicated by the vertical line. In comparison with the equilibrium 
values we can see that the variance and kurtosis, which have a simple structure as a function of temperature (see 
Fig.~\ref{fig:integratedmoments}), 
have smaller extremal values, which are about 75\% of the equilibrium variance and about 50\% of the equilibrium kurtosis. 
This is an effect of the long relaxation times for modes which are of the order of the inverse correlation length, see 
Fig.~\ref{fig:RelaxationTimeScaling}, and the non-equilibrium situation of a rapidly cooling system. In addition we can observe a 
dynamical retardation effect, which shifts these extrema (a slightly stronger shift is observed for $(\kappa\sigma^2)_V$) to times 
larger than $\tau_c$. The skewness follows qualitatively this behavior, but quantitative statements are difficult 
given the larger uncertainties. The observed effects, the decrease of extremal values and the retardation effect, are found to be 
stronger for slower diffusion and/or faster cooling. 

\section{Conclusions}

In this work we presented first results of a fully dynamical treatment of the diffusive behavior of fluctuations near the 
QCD critical point. Our study takes into account nonlinear mode couplings, the finite size of the system, and exact net-baryon 
number conservation. In equilibrium we find that these effects limit the growth of the correlation length $\xi$ near $T_c$. We 
observe that the nonlinear couplings generate non-Gaussian fluctuations from the Gaussian stochastic noise. 
The scaling behavior of the variance and the higher-order cumulants with the correlation length is affected by a non-trivial interplay 
between nonlinear mode couplings, the finite ratio of $\xi$ over the observation volume $V$, and the impact of net-baryon number 
conservation. These effects can be studied systematically as an expansion in $\xi/V$. 

From the scaling behavior of the relaxation time of the critical modes we demonstrate that the dynamical scaling of model $B$ is 
realized in the presented framework. In the dynamics of the fluctuations we see clear signatures of retardation and 
non-equilibrium effects. Critical slowing down leads to a visible reduction of the magnitude of the fluctuations, which is stronger 
for the higher-order cumulants. Our work marks an important step in the theoretical development of dynamical models guiding the 
search for the QCD critical point. 

\section*{Acknowledgments}
M. Nahrgang acknowledges the support of the program ``Etoiles montantes en Pays de la Loire 2017''. The work of M. Bluhm is funded 
by the European Union’s Horizon 2020 research and innovation program under the Marie Sk\l{}odowska Curie grant agreement No 665778 
via the National Science Center, Poland, under grant Polonez UMO-2016/21/P/ST2/04035. T. Sch\"afer was supported in part by the U.S. 
Department of Energy under grant DE-FG02-03ER41260 and by the DoE Beam Energy Scan Theory (BEST) Topical Collaboration. S.A. Bass 
has been supported by the U.S. Department of Energy under grant DE-FG02-05ER41367. This research was supported in part by the 
ExtreMe Matter Institute (EMMI) at the GSI Helmholtzzentrum f\"ur Schwerionenforschung, Darmstadt, Germany. The authors acknowledge 
fruitful discussions within the framework of the BEST Topical Collaboration.

\end{document}